\documentclass[seceq]{ptptex}

\usepackage{graphicx}
\usepackage{amsmath}

\title{Critical properties of phase transitions\\
in lattices of coupled logistic maps}

\author{Philippe \textsc{Marcq}$^1$, Hugues \textsc{Chat\'e}$^2$, and
Paul \textsc{Manneville}$^3$}

\inst{
$^1$ Institut de Recherche sur les Ph\'enom\`enes Hors \'Equilibre,\\
49 rue Joliot-Curie, BP 146, 13384 Marseille Cedex 13, France\\
$^2$ CEA --- Service de Physique de l'\'Etat Condens\'e,\\
Centre d'\'Etudes de Saclay, 91191 Gif-sur-Yvette, France\\
$^3$ LadHyX -- Laboratoire d'Hydrodynamique,\\ \'Ecole Polytechnique,
91128 Palaiseau, France
}

\abst{We numerically demonstrate that collective bifurcations in 
two-dimensional lattices of locally coupled logistic maps share most 
of the defining features of equilibrium second-order phase transitions.
Our simulations suggest that these transitions between distinct 
collective dynamical regimes belong 
to the universality class of Miller and Huse model with synchronous update
[Marcq et al., Phys. Rev. Lett. {\bf 77} (1996), 4003].
}

\begin{document}

\maketitle

\section{Introduction}
\label{sec:itd}

For strong enough coupling, mutually interacting oscillators tend to synchronize.
Prof. Kuramoto first gave a solid mathematical grounding to this rather intuitive idea 
in 1975 \cite{Kuramoto}. More precisely, he showed analytically that an ensemble of 
nonidentical phase oscillators with distributed natural frequencies, and coupled through 
their mean-field, synchronize above a critical value of the coupling constant 
\cite{BookKuramoto}. This dynamical phase transition is a robust phenomenon, 
generically occurring in  a wide class of similar coupled dynamical systems \cite{RMPKuramoto}.

The system of interest in the present work is complementary to Kuramoto's model: 
$N$ identical units with
aperiodic individual (discrete time) dynamics sit on the nodes of a regular lattice,
and are updated synchronously with local (diffusive) coupling \cite{Kaneko}. 
For a large enough coupling constant, the spatially-averaged activity of such
coupled map lattices is time-periodic \cite{ChateManneville}. 
However, synchronization of individual units is \emph{not} involved, at least in the
usual sense: macroscopic coherence coexists with microscopic disorder, as evidenced
by broad distribution functions of local activity and rapidly
decaying spatial correlation functions. Extensive numerical simulations
have confirmed that these emergent macroscopic cycles
are global attractors, well-defined in the infinite-time, infinite-size 
(thermodynamic) limit \cite{ChateManneville}: fluctuations 
of the spatially-averaged activity about the collective cycle vanish when $N \rightarrow \infty$. 

For a large enough control parameter, a single logistic map exhibits an
inverse bifurcation cascade between regimes of banded chaos. In the same
parameter region, and for strong enough coupling, locally coupled
logistic maps exhibit an inverse bifurcation cascade between macroscopic cycles 
of period $2^n$ \cite{ChateManneville}. Generalized mean-field arguments 
give a satisfactory understanding of the build-up of correlations at the 
origin of the dynamical long-range order involved here \cite{AnaelMeanField}, at least far
from bifurcation points. 

In this work, we wish to characterize, thanks to numerical simulations, the first bifurcation points
in the cascade \cite{These}. Upon defining adequate order parameters, we demonstrate
that these macroscopic bifurcations harbor the characteristic 
features of \emph{equilibrium} second-order phase transitions. The relevance of
standard finite-size scaling theory indicates that they persist in the infinite-size limit.
Many properties of an equilibrium system close to a second-order transition 
turn out to be largely independent of the microscopic details of the interactions between individual
components: they fall instead into a small number of universality classes,
each defined by global features such as the symmetries of the underlying
Hamiltonian or the spatial dimensionnality of the system. 
We numerically evaluate the critical exponents of period-doubling 
phase transitions, and discuss the relevance of the notion of universality
to \emph{temporal} spontaneous symmetry breaking.

\section{Critical properties of a period-doubling phase transition}
\label{sec:p1p2}

We consider the dynamics of a set of $N = L^2$ variables $x_{i,j}^t$
sitting on the nodes of a two-dimensional square lattice.
Time is discrete. The update rule consists of two stages: each site 
is first updated according to the logistic map with parameter $r \in [0,4]$:
\begin{equation}
  \label{eq:logistic}
  \begin{array}{lcl}
f : &[0,1] &\rightarrow [0,1],\\
    & x &\mapsto r \; x \; (1-x),
\end{array}
\end{equation}
then transformed according to a diffusive coupling operator. 
Interaction is restricted to nearest-neighbors.
All sites are updated \emph{synchronously} according to the rule:
\begin{equation}
  \label{eq:evol}
  x_{i,j}^{t+1} = (1-4 g)\; f(x_{i,j}^t) + g \; \left(
f(x_{i-1,j}^{t}) +  f(x_{i,j-1}^{t}) +  f(x_{i+1,j}^t)
+  f(x_{i,j+1}^t) \right),
\end{equation}
where $g$ is the coupling constant, set to $g = 0.2$
in this Section (democratic coupling).
The \emph{microscopic} control parameter of the lattice dynamical system is $r$.

\begin{figure}[t]
\centerline{\includegraphics[width=6cm,height=5cm]{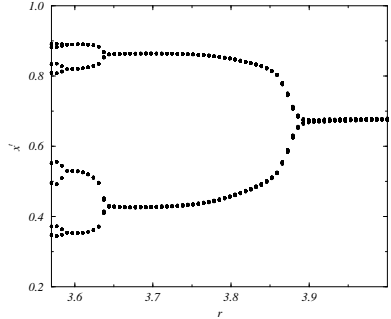}}
\caption{\label{fig:diagbifg0.2}
Bifurcation diagram of democratically coupled ($g = 0.2$) logistic maps
on a two-dimensional lattice of $L^2 = 1024^2$ sites: for each parameter value 
$r \ge r_{\infty}$, $20$ consecutive values of the mean activity $x^t$ are plotted vs. $r$. 
Initial site values are randomly distributed over the interval $[0,1]$, 
and a transient of duration $t_0 = 10^4$ is discarded. Period-doubling bifurcations
are rounded by finite-time, finite-size effects.}
\end{figure}

\begin{figure}[th]
\hfill
\includegraphics[width=6cm,height=5cm]{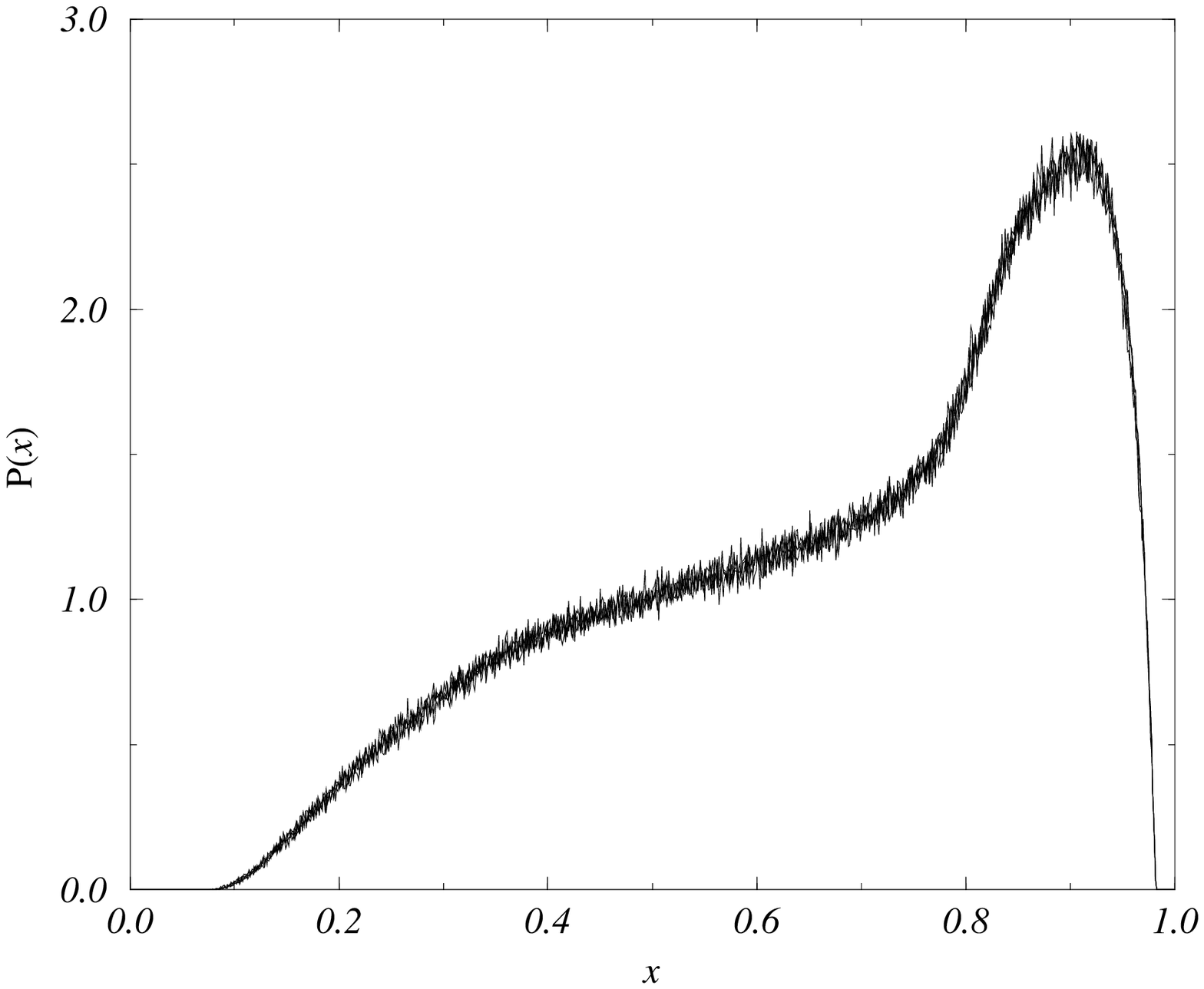}
\hfill
\includegraphics[width=6cm,height=5cm]{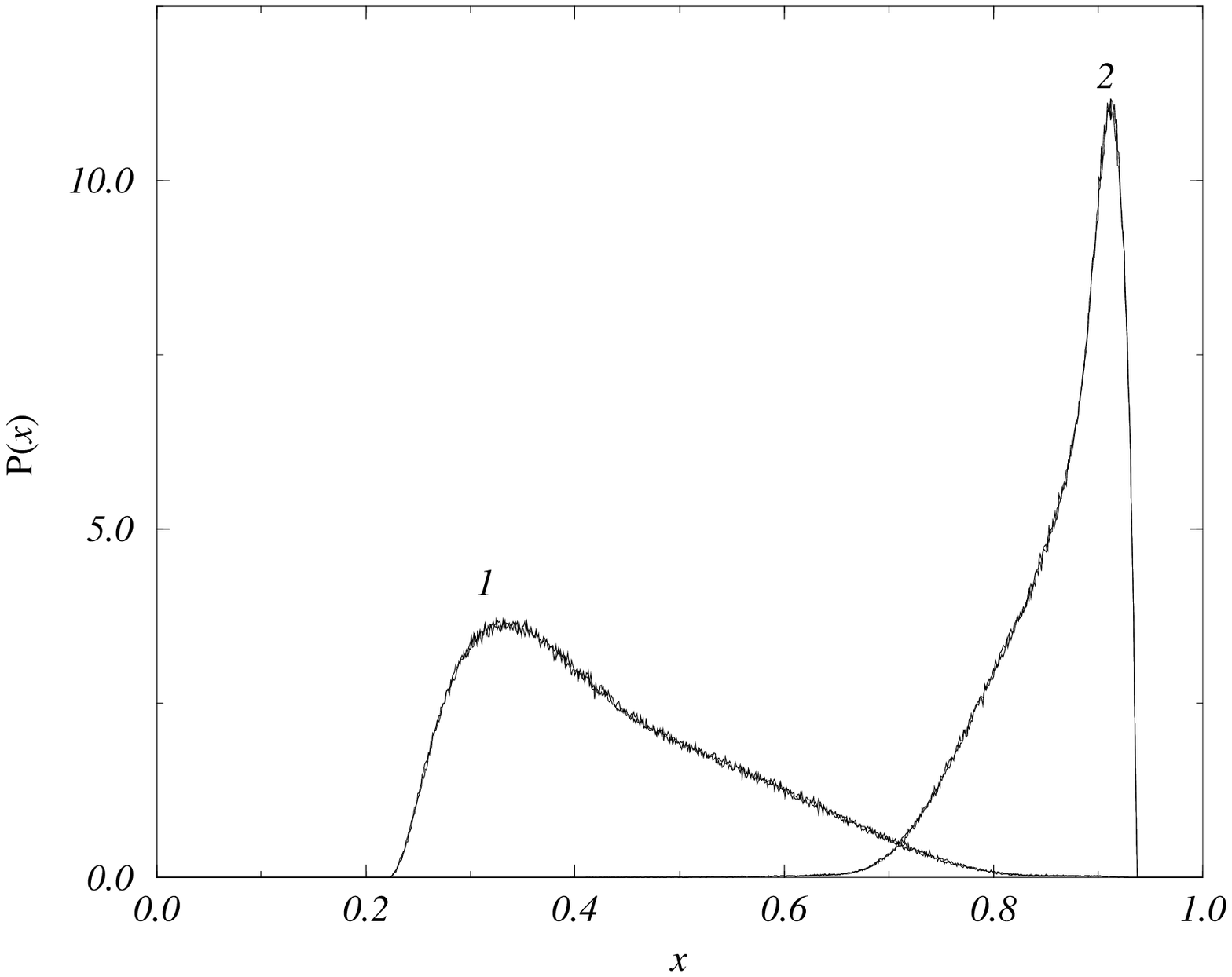}
\hfill
\caption{\label{fig:pdfp}                                   
Probability distribution function of the lattice's site values, measured 
for $g = 0.2$,  $L = 1024$, in the period 1 (left-hand graph, $r = 3.93$)
and period 2 (right-hand graph, $r = 3.75$) collective states. The pdfs 
measured over $4$ consecutive time-steps are superposed. The breadth of the lines
is related to finite-size statistical fluctuations.}
\end{figure}

The bifurcation diagram of the mean activity
\begin{equation}
  \label{eq:meanfield}
  x^t  =  {1 \over L^2} \sum_{i,j=1}^L x_{i,j}^t
\end{equation}
is given in Fig.~\ref{fig:diagbifg0.2} for $r > r_\infty = 3.57\ldots$.
The statistical behavior of the coupled map lattice may be interpreted as long-range 
order accompanied by the temporal evolution of spatially-averaged quantities:
fixed point (or period 1) above $r_{1-2} \simeq 3.86$, period 2 for 
$r \in [r_{2-4}, r_{1-2}]$, with $r_{2-4} \simeq 3.63$, period $2^n$, $n \ge 2$,
for $r \in [r_{\infty}, r_{2-4}]$. An infinite cascade of period-doubling bifurcations
is expected to occur in the limit $r \rightarrow r_\infty$ \cite{AnaelRenormalization}.

For simplicity, we focus on the period 1-period 2 bifurcation.
The (time-asymptotic) distribution of site values in these two regimes
is shown for typical parameters in Fig.~\ref{fig:pdfp}.
The Lyapunov spectrum scales linearly with the system size: chaos is
\emph{extensive}. Dynamical quantifiers such as the largest Lyapunov exponent and 
the Kolmogorov entropy remain continuous close to the bifurcation point (see also \cite{Greenside}).

The local ``magnetization'' is defined by $m_{1-2}^t = x^{2t+1} - x^{2t}$. 
The order parameter of the transition reads \cite{These,WLH}:
\begin{equation}
  \label{eq:paramp1p2}
  M_{1-2} = \left\langle \,\left| m_{1-2}^t \right|\, \right\rangle =
\lim_{T \rightarrow \infty} {1 \over T} \sum_{t = 0}^T 
\left| x^{2t+1} - x^{2t} \right|.
\end{equation}
The period 1 and period 2 phases are respectively ``paramagnetic'' ($M_{1-2} = 0$)
and ``ferromagnetic'' ($M_{1-2} \neq 0$). Through Eq.~(\ref{eq:paramp1p2}),
we wish to draw a strong analogy with the Ising model. 
Instead of spin-reversal invariance, time-translation invariance is broken in the ordered phase
of the lattice dynamical system.
Further, the correlation length $\xi$ exhibits a sharp maximum close to 
the transition, where $\xi$ characterizes the exponential decay of
the equal-time, two-point correlation function of the field $x_{i,j}^t$.
Large scale simulations suggest the presence of well-defined 
power laws controling the behavior close to the transition of 
the correlation length $\xi$, the magnetization 
$M_{1-2}$, and the susceptibility $\chi_{1-2}$, defined by 
$\chi_{1-2} = L^2 \; \left\langle \; (|m_{1-2}^t| - M_{1-2})^2 \; \right\rangle$.

\begin{figure}[t]
\hfill
\includegraphics[width=6cm,height=5cm]{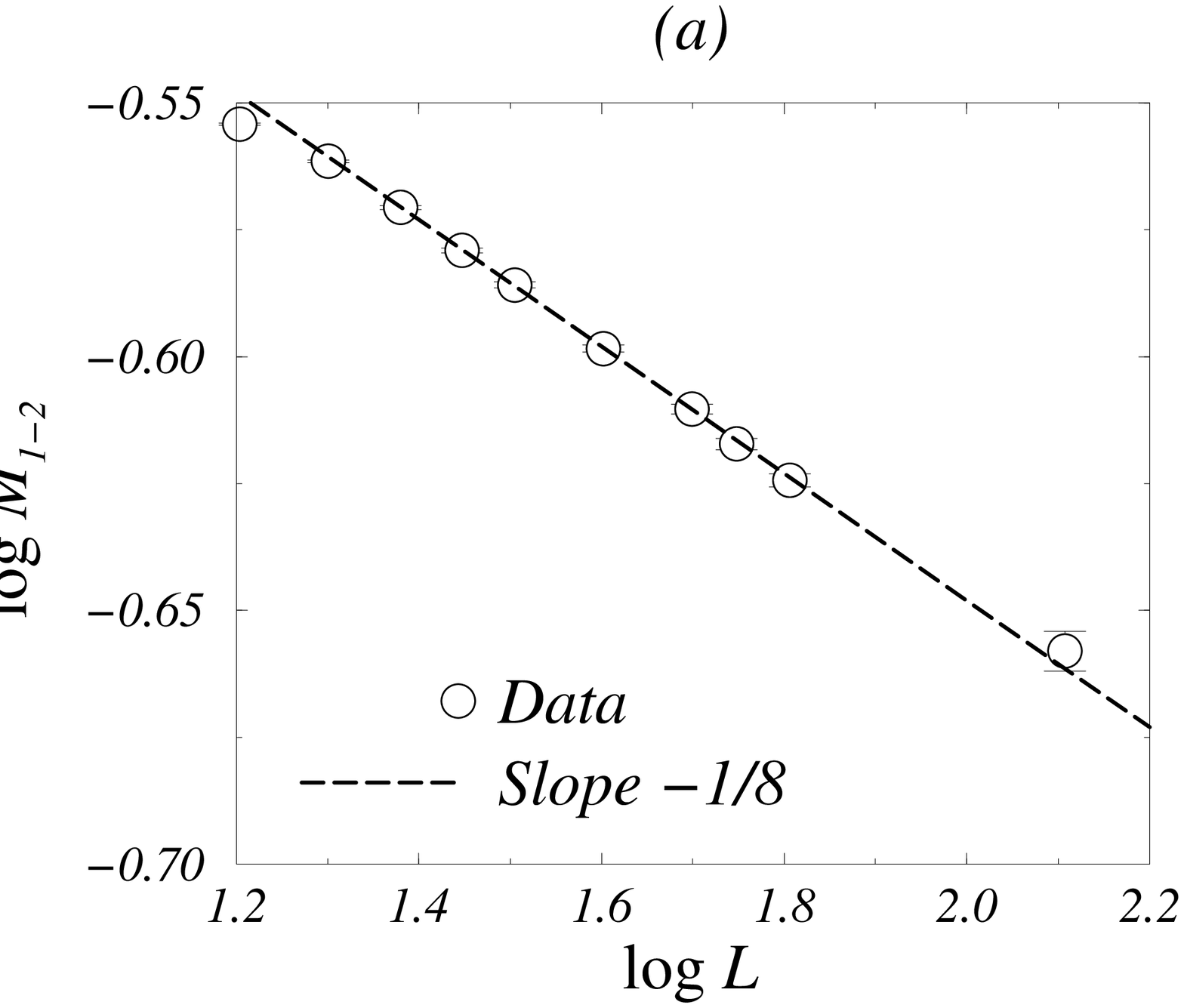}
\hfill
\includegraphics[width=6cm,height=5cm]{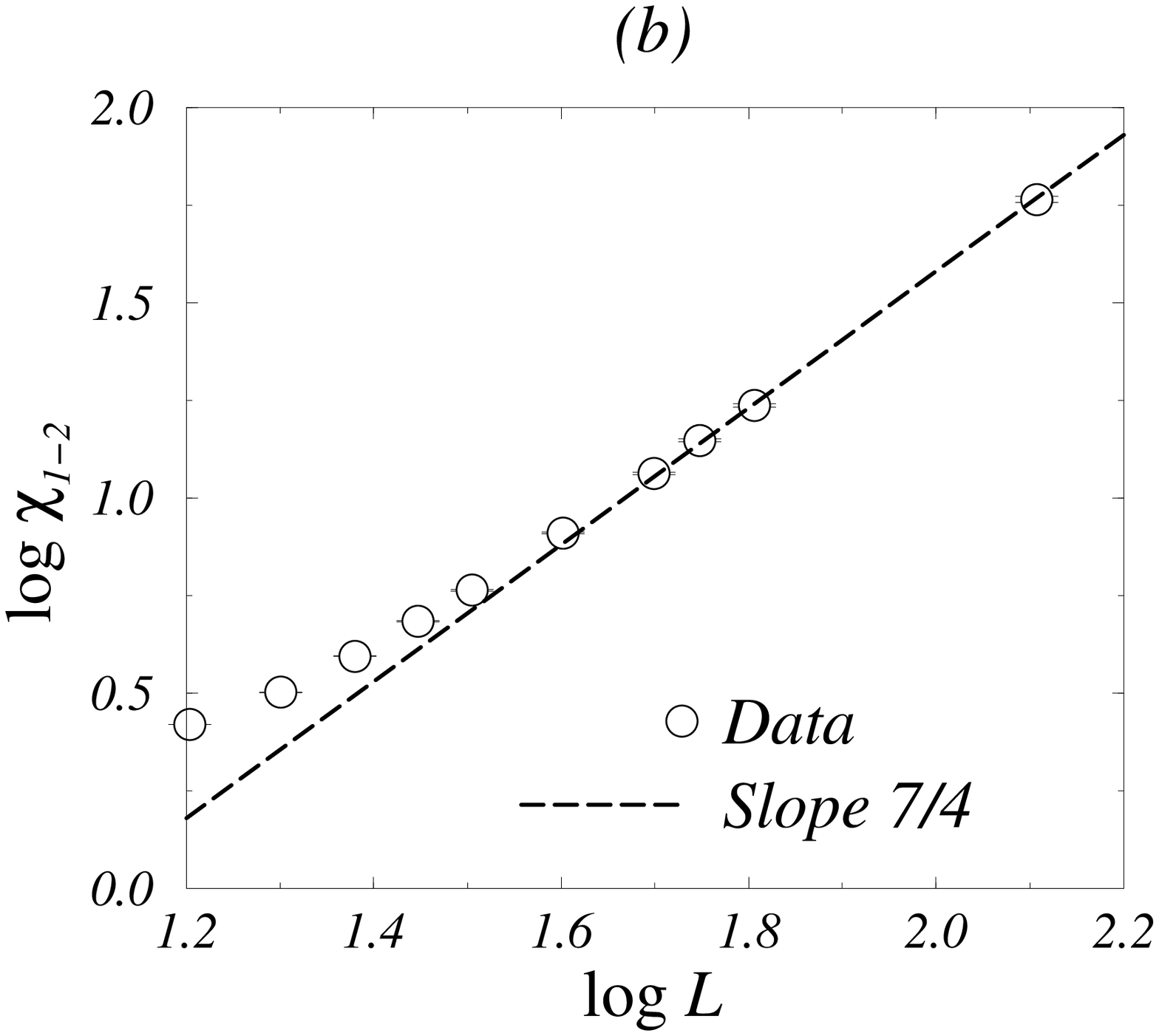}
\hfill
\centerline{\includegraphics[width=6cm,height=5cm]{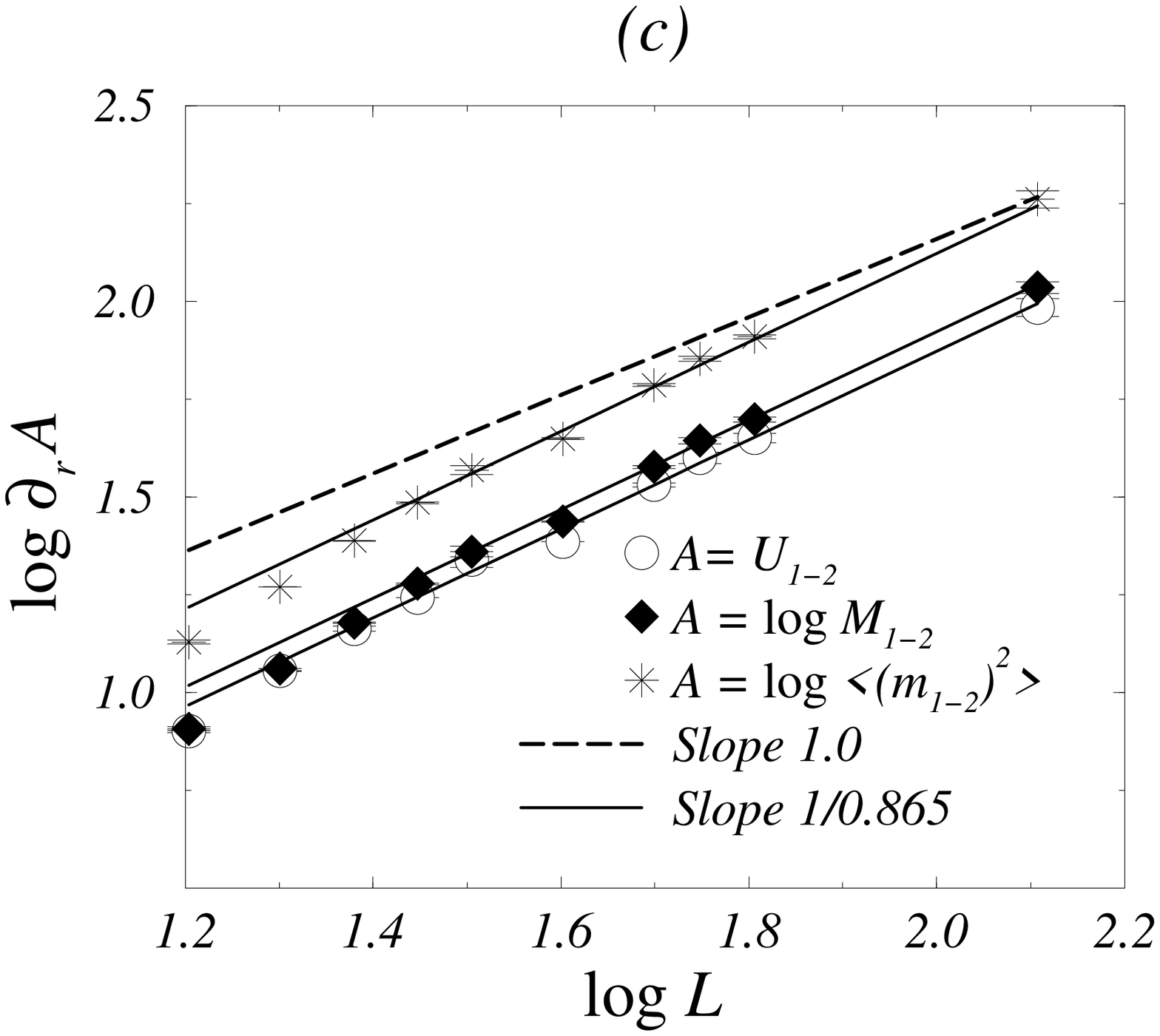}}
\caption{\label{fig:p1p2}                                   
Period 1-period 2 transition, $g = 0.2$, $r =r_{1-2}^{\infty} = 3.86212$.
The finite-size scaling laws given in Eq.~(\ref{eq:scale}) allow to measure the
exponent ratios $\beta/\nu$ (a), $\gamma/\nu$ (b), $1/\nu$ (c). Corrections to scaling
are apparent in the three graphs. The dashed lines correspond to the exact exponent values 
of the two-dimensional Ising model.}
\end{figure}

The precise location of the transition point is determined
using Binder's method \cite{Binder}: the value of the finite-size cumulant 
$U_{1-2}^L(r) = -3+ \left\langle (m_{1-2}^t)^4 \right\rangle/
 \left\langle (m_{1-2}^t)^2 \right\rangle^2 $
at size $N = L^2$ is independent of $L$ at criticality: $\forall L$,
$U_{1-2}^L(r_{1-2}^{\infty}) = U_{1-2}^{\infty}$.
Using system sizes ranging from $L = 32$ to $L = 128$,
we find $r_{1-2}^{\infty} = 3.86212(12)$, 
$- U_{1-2}^{\infty} = 1.819(12)$, where numbers within parentheses
correspond to the uncertainty over the last digit(s).
The exponent ratios $\beta/\nu$, $\gamma/\nu$, $1/\nu$
are then obtained from the following scaling laws:
\begin{equation}
\label{eq:scale}
\begin{array}{llll}
M_{1-2}(L) &\propto& L^{-\beta/\nu}  ,\\
\chi_{1-2}(L) &\propto& L^{\gamma/\nu} ,\\
\partial_r U_{1-2}^L &\propto& L^{1/\nu} ,\\
\partial_r \log M_{1-2}(L) &\propto& L^{1/\nu},\\
\partial_r \log \left\langle m_{1-2}^t(L)^2 \right\rangle &\propto& L^{1/\nu},
\end{array}
\end{equation}
where statistical averages are computed at the infinite-size critical 
parameter value $r_{1-2}^{\infty}$.
For all the above observables ${\mathcal O}$, our data is consistent with the 
simplest equilibrium correction to scaling:
${\mathcal O} = L^{\phi_{\mathcal O}}   \left(a_0 + a_1 L^{-\omega_{\mathcal O}} + \ldots\right)$,
where $\phi_{\mathcal O}$ and $\omega_{\mathcal O}$ are respectively the 
scaling exponent defined in Eqs.~(\ref{eq:scale}) and 
the first subdominant exponent.
Taking into account this correction, our best estimates are (see \cite{IsingCML} for
a detailed account of the procedure we use):
\begin{equation}
  \label{eq:valexpp1p2}
\begin{array}{llll}
\beta/\nu & = & 0.126(4),\\
\gamma/\nu & = & 1.76(3),\\
\nu &=& 0.865(25).
\end{array}
\end{equation}

The exponent ratios $\beta/\nu$ and $\gamma/\nu$ are in excellent agreement with 
the two-dimensional equilibrium Ising model:
$(\beta/\nu)_{\mathrm{Ising}} = 1/8$, $(\gamma/\nu)_{\mathrm{Ising}} = 7/4$.
However, the correlation length exponent $\nu$ is \emph{not} consistent with 
$\nu_{\mathrm{Ising}} = 1$ (see Fig.~\ref{fig:p1p2}).
As a consequence, our estimates for $\beta = 0.108(6)$ and $\gamma = 1.52(7)$ 
are also incompatible with the Ising values. Unless corrections to scaling of
an unusual nature are present in this system, our data suggests that 
the period 1-period 2 \emph{phase transition} does not belong to the
universality class of the two-dimensional equilibrium Ising model:
the analogy we have drawn breaks down at a quantitative level.

\section{Universality}
\label{sec:uni}

The evolution rule (\ref{eq:logistic})-(\ref{eq:evol}) involves two parameters: 
the coupling constant $g$ and the nonlinear parameter $r$:
a line of period 1-period 2 transitions is observed in the parameter plane
for large enough coupling ($g \gtrsim 0.10$).
We studied two other transition points on this line: a $r$-driven transition
at $g = 0.11$, and a coupling-driven transition at a fixed value of $r = 3.831$.
The same measurement protocol is used, with similar system sizes and statistical
accuracy (see \cite{These} for further details). In the three cases, we find exponent
values mutually consistent within error bars (see Table for a summary of results). 
This suggests that the period 1-period 2 transition line defines a universality class 
distinct from that of the two-dimensional Ising model. However, the hyperscaling 
relation $2 \beta + \gamma = d \nu$ ($d = 2$) remains valid, and the critical value
$U^{\infty}$ of Binder's cumulant is consistent with the Ising value.

We also studied transitions between cycles of higher period, which
are characterized by the same phenomenology as described above.
Physical observables are defined as before, upon replacing 
$m^t_{1-2}$ by the appropriate local magnetization, for instance
$m^t_{2-4} = x^{4t+2} - x^{4t}$ for period 2-period 4 transitions.
We find that: $(i)$ these transitions do not belong to the Ising
universality class; $(ii)$ the measured exponent values are
consistent with those found for period 1-period 2 transitions --
even though the system sizes we use do not allow a clear-cut
answer to this last question.

\begin{table}[t]
\label{table}
\begin{center}
\begin{tabular}{l|c|c|ccc}
\hline\hline
                        & 2D Ising & MH & $g = 0.2$ & $g = 0.11$ & $r = 3.831$ \\ \hline
Threshold               & $$  & $0.20534(2)$  & $3.86212(12)$ & $3.8310(1)$ & $0.1100(1)$ \\
$-U^{\infty}$                  & $1.83$    & $1.832(4)$    & $1.819(12)$   & $1.828(10)$   & $1.83(3)$   \\ \hline
$\beta/\nu$             & $0.125$    & $0.125(4)$    & $0.126(4)$    & $0.131(10)$    & $0.129(10)$      \\
$\gamma/\nu$            & $1.75$   & $1.748(10)$   & $1.76(3)$     & $1.75(4)$     & $1.73(3)$         \\
$(2 \beta+\gamma)/\nu$  & $2$     & $2.00(2)$     & $2.02(4)$     & $2.01(6)$     & $1.99(5)$       \\ \hline
$\beta$                 & $0.125$    & $0.111(5)$    & $0.108(6)$    & $0.107(21)$    & $0.107(10)$     \\
$\gamma$                & $1.75$     & $1.55(4)$     & $1.52(7)$     & $1.50(12)$     & $1.39(10)$      \\
$\nu$                   & $1$   & $0.887(18)$   & $0.865(25)$   & $0.860(45)$   & $0.80(4)$   \\ 
\hline\hline
\end{tabular}
\end{center}
\caption{Summary of critical exponents: Ising model; 
Miller and Huse model; $r$-driven period 1-period 2 transition at $g = 0.2$;
$r$-driven period 1-period 2 transition at $g = 0.11$;
$g$-driven period 1-period 2 transition at $r = 3.831$.}
\end{table}

In fact, the same features are characteristic of other ordering transitions between chaotic
phases. Miller and Huse introduced a two-dimensional coupled map lattice with a 
microscopic ``up-down'' symmetry, using a continuous, piecewise-linear, odd-symmetric 
map  \cite{MillerHuse}. In a previous work  \cite{IsingCML}, we
showed that the Ising-like transition of this extensively chaotic lattice dynamical system
does not belong to the Ising universality class, with in particular 
$\nu = 0.89 \pm 0.03$. However, Ising exponents are recovered 
for the same geometry, coupling, and local map once the update rule becomes asynchronous 
(lattice sites updated one at a time). The nature of update, a dynamical feature, 
is a relevant ``parameter'' that distinguishes between universality classes. 
The critical exponents of Miller and Huse's model with synchronous update are recalled 
in the Table: we showed \cite{IsingCML} that they are representative
of a genuine universality class in the sense that changes of, e.g., the local map
do not alter them provided that the up-down symmetry is preserved
(see also \cite{SastrePerez,Makowiec}). Remarkably, these exponents are consistent 
(within error bars) with those obtained for the period 1-period 2 transitions 
of two-dimensional coupled logistic maps. The same universality class
encompasses ordering transitions between chaotic phases of synchronously-updated
lattice dynamical systems, whether due to the (discrete) breaking of 
an up-down symmetry in phase space or to that of time-translation invariance.

\section{Conclusion}
\label{sec:ccl}

Our numerical simulations demonstrate that \emph{equilibrium} finite-size 
scaling laws allow to characterize the period-doubling macroscopic bifurcations
of lattices of locally coupled chaotic logistic maps. This relevance provides in itself 
further evidence that such bifurcations are indeed continuous phase transitions,
well-defined in the infinite-size limit, where the equal-time correlation length 
diverges as the system's linear size. We evaluate 
the static critical exponents $\beta$, $\gamma$ and $\nu$ of fixed point-period 2 
and period 2-period 4 transitions. The best interpretation of our (finite-size) data is the following:
these transitions belong to the universality class of the Miller and
Huse model with synchronous update, \emph{not} to the
equilibrium Ising universality class: $\nu \simeq 0.89 \neq \nu_{\mathrm{Ising}} = 1$.

Period-doubling phase transitions are by all means unusual: they separate
microscopically chaotic states with different (regular) macroscopic \emph{dynamics}.
The existence of an ordered phase is related to the discrete breaking
of time-translation invariance at the macro-scale, while microscopic dynamical quantifiers,
such as the Lyapunov spectrum, remain continuous close to the transition.
The lattice dynamical system is a priori far from equilibrium: 
microreversibility is not expected to hold.
The non-Ising value of the correlation length exponent $\nu$, at odds
with standard coarse-graining arguments \cite{MillerHuse,Egolf}, 
remains a puzzle.

Two comments are in order. First, the control parameters
we use are \emph{microscopic}. This is also true of related models with
similar critical properties \cite{SastrePerez,Makowiec,Lyapunov}.
The various phases we observe arise due to a balance between nearest-neighbor
interactions and fluctuations of deterministic origin. Since the critical
exponents we measure are not those of mean-field theory \cite{MeanField},
we know that fluctuations cannot be neglected close to the transition points.
One would certainly like to be able to satisfactorily define the temperature
of a given extensively chaotic lattice dynamical system, \emph{i.e.} 
a macroscopic, intensive parameter, well-defined in the infinite-size limit, that quantifies
at a coarse-grained level the degree of microscopic disorder. 
This remains an open, challenging problem.

Second, the synchronous nature of the update rule of a coupled map lattice is never
insignificant \cite{update1,update2}. In the case of Miller and Huse model, 
Ising exponents are recovered with asynchronous update: the nature of update 
is a relevant ``parameter'' in the sense of critical phenomena. For coupled logistic maps, 
the long-range order leading to periodic collective behavior is destroyed by 
asynchronous update. The fully synchronized state, where all lattice site values are equal 
to that of the unstable fixed point of the local map, remains unstable under synchronous update. 
However, synchronization transitions occur as soon as the update rule 
includes some degree of asynchrony \cite{Sinha}.
In all cases, the choice of a given update rule is a crucial modeling issue.

\end{document}